\documentclass[sigconf]{acmart}
\AtBeginDocument{%
  \providecommand\BibTeX{{%
    \normalfont B\kern-0.5em{\scshape i\kern-0.25em b}\kern-0.8em\TeX}}}

\copyrightyear{2025}
\acmYear{2025}
\setcopyright{cc}
\setcctype{by}
\acmConference[CHI '25]{CHI Conference on Human Factors in Computing Systems}{April 26-May 1, 2025}{Yokohama, Japan}
\acmBooktitle{CHI Conference on Human Factors in Computing Systems (CHI '25), April 26-May 1, 2025, Yokohama, Japan}\acmDOI{10.1145/3706598.3713128}
\acmISBN{979-8-4007-1394-1/25/04}

\acmSubmissionID{9728}

\usepackage{color}
\usepackage{colortbl}
\usepackage{xcolor}
\usepackage{makecell}
\usepackage{multirow}
\usepackage{multicol}
\usepackage{xspace}
\usepackage{hyperref}
\newcommand{\name}
{\textsc{Chartist}\xspace}
\usepackage{subfigure}
\usepackage{pifont}
\usepackage{xcolor}

\usepackage{ifthen}
\newboolean{revising}
\setboolean{revising}{false}
\ifthenelse{\boolean{revising}}
{
    \newcommand{\rv}[1]{\textcolor{blue}{#1}}
} {
    \newcommand{\rv}[1]{#1}
}

\begin{document}
\title{\rv{\name: Task-driven Eye Movement Control for Chart Reading}}
% \title{How People Read Charts: A Model of Task-driven Eye Movement Control}
% \title{Modeling Task-driven Scanpaths on Charts}
% \title{Task-driven Human Attention Prediction for Chart Question Answering}


\author{Danqing Shi}
\orcid{0000-0002-8105-0944}
\affiliation{%
  \institution{Aalto University}
  \state{Helsinki}
  \country{Finland}}
\author{Yao Wang}
\orcid{0000-0002-3633-8623}
\affiliation{%
  \institution{University of Stuttgart}
  \state{Stuttgart}
  \country{Germany}}
\author{Yunpeng Bai}
\orcid{0009-0008-7578-0079}
\affiliation{%
  \institution{National University of Singapore}
  \state{Singapore}
  \country{Singapore}}
\author{Andreas Bulling}
\orcid{0000-0001-6317-7303}
\affiliation{%
  \institution{University of Stuttgart}
  \state{Stuttgart}
  \country{Germany}}
\author{Antti Oulasvirta}
\orcid{0000-0002-2498-7837}
\affiliation{%
  \institution{Aalto University}
  \state{Helsinki}
  \country{Finland}}

\renewcommand{\shortauthors}{Shi et al.}

%   \centering
%   \includegraphics[width=0.9\textwidth]{Images/teaser-scanpath.png}
%   \caption{We present \name, a computational model that can predict task-driven human scanpaths on charts. The figure demonstrates three analytical tasks involved in the study: retrieve value, filter, and find extreme. The visualization illustrates how models' predictions vary across tasks and match the pattern of human scanpaths, with fixation density maps overlaid.}

% \begin{figure}[b]
% \noindent\fbox{%
% \parbox{\dimexpr\linewidth-2\fboxsep-2\fboxrule\relax}{%
% \begin{tabular}{l}
% The word count of this paper is \textcolor{blue}{7357}.
% \end{tabular}
% }%
% }
% \end{figure}

%% article

\begin{abstract}
To design data visualizations that are easy to comprehend, we need to understand how people with different interests read them. Computational models of predicting scanpaths on charts could complement empirical studies by offering estimates of user performance inexpensively; however, previous models have been limited to gaze patterns and overlooked the effects of tasks. Here, we contribute \name, a computational model that simulates how users move their eyes to extract information from the chart in order to perform analysis tasks, including value retrieval, filtering, and finding extremes. 
The novel contribution lies in a two-level hierarchical control architecture. At the high level, the model uses LLMs to comprehend the information gained so far and applies this representation to select a goal for the lower-level controllers, which, in turn, move the eyes in accordance with a sampling policy learned via reinforcement learning.
\rv{
The model is capable of predicting human-like task-driven scanpaths across various tasks. It can be applied in fields such as explainable AI,  visualization design evaluation, and optimization. 
While it displays limitations in terms of generalizability and accuracy, it takes modeling in a promising direction, toward understanding human behaviors in interacting with charts.
}
\end{abstract}


\maketitle

\vspace{-0.5mm}
\section{Introduction}

\begin{figure*}[!t]
  \centering
  \includegraphics[width=\textwidth]{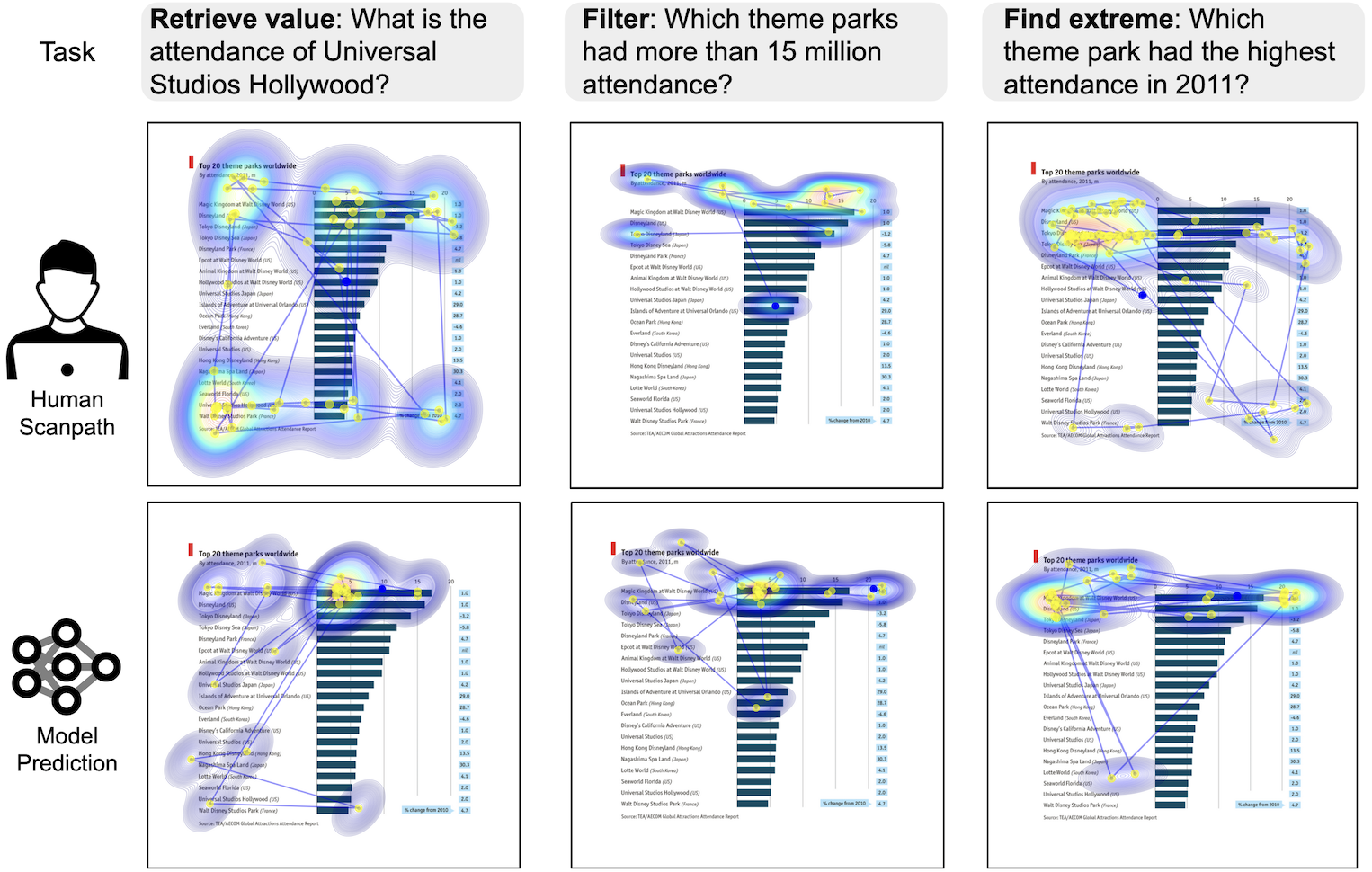}
  \caption{We present \name, a computational model that can predict task-driven human scanpaths on charts. The figure demonstrates three analytical tasks involved in the study: retrieve value, filter, and find extreme. The visualization illustrates how models' predictions vary across tasks and match the pattern of human scanpaths, with fixation density maps overlaid.}
  \label{fig:teaser}
\end{figure*}

Visual attention plays a pivotal role in the field of information visualization~\cite{healey2011attention, borkin2015beyond}.
By understanding the visual attention on charts, designers can iteratively refine visualizations using visual saliency as feedback~\cite{shin2023perceptual}; engineers can enhance their AI models by incorporating human-like attention ~\cite{sood2023multimodal}; and researchers can better understand the link between comprehension and gaze behavior when people read charts~\cite{wang2024visrecall++}.
Eye tracking has long been used to understand human visual attention on charts~\cite{goldberg2011eye}.
Beyond visual saliency~\cite{shin2022scanner}, analyzing human scanpaths provides details of the sequence of fixations, helping researchers understand strategies for reasoning~\cite{wang2023scanpath}.
Previous literature has demonstrated that users observe completely different visual elements when performing different analytical tasks~\cite{polatsek2018exploring}.
However, collecting human scanpaths by using eye trackers is expensive both time-wise and monetarily.
\rv{   
Simulations are effective in developing theories by rigorously testing user interactions with visual elements in controlled settings. By simulating eye movements, researchers can uncover the mechanisms behind behaviors of users interpreting data visualizations. This enhances understanding of chart reading~\cite{murray2022simulation}.
}

Human visual attention is guided by two processes: bottom-up and top-down processes~\cite{itti2000saliency}. These processes apply to chart reading as well~\cite{yang2024unifying}. Bottom-up attention is driven by salient visual stimuli (e.g., high-contrast colors), whereas top-down attention is task-driven, with specific goals or intentions shaping where and how users focus their attention.
However, most visual attention models applied for information visualizations capture only bottom-up (free-viewing) attention~\cite{matzen2017data, shin2022scanner, wang2023scanpath}, thus leaving a gap in understanding how tasks influence human visual attention~\cite{amar2005low}.
Compared to exploratory free viewing, scanpaths for the same analysis task are more coherent; also, they vary greatly between tasks~\cite{polatsek2018exploring}.

While recent research has been able to predict task-driven saliency on charts~\cite{wang2024salchartqa}, it has remained one step away from addressing how people read charts. 
Temporal information and individual-specific behaviors are still missing from task-driven-saliency maps.
In other words, what would the scanpath look like when a person carries out a particular task on a given chart?
In this paper, we present \name, the first computational model for predicting task-driven scanpaths on charts~\footnote{\href{https://chart-reading.github.io}{https://chart-reading.github.io}}.
When given both an image of a chart and a sentence as the analytical task, \name can simulate a sequence of human-like fixation positions related to the task (see Figure~\ref{fig:teaser}).
The model has two key distinctions from preexisting models for scanpath prediction: 
1) Our model focuses on predicting fixations made during analytical tasks, including both fixation positions and their order, in contrast against the current state-of-the-art models, which concentrate on free-viewing conditions. Task factors' influence makes it challenging to predict task-driven scanpaths via prior models.
2) Unlike goal-driven scanpath predictions, which are limited to visual search tasks with natural images~\cite{mondal2023gazeformer}, analytical tasks require high-level reasoning and also tackling the limitations of human cognition in chart question answering.

\begin{figure*}[!t]
\centering
  \includegraphics[width=0.9\textwidth]{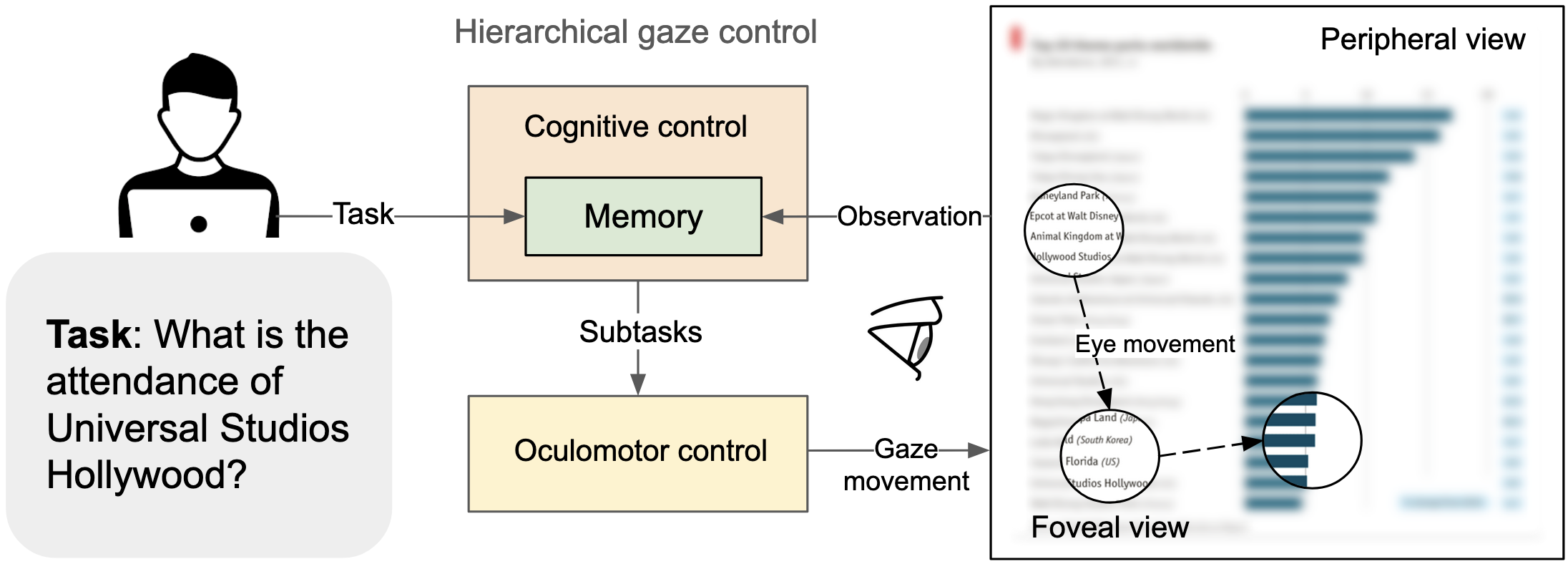}
  \caption{The figure illustrates the concept of the model for task-driven eye movement control. When given a task, the agent makes decisions about the next subtask, based on information gathered from observing the chart stored in its memory. Each subtask controls eye movements at pixel level and retrieves information from the foveal vision area of the gaze.}
  \Description{Concept of the model for task-driven eye movement control}
  \label{fig:concept}
\end{figure*}

To enable such capabilities, we propose a hierarchical control architecture for modeling (see Figure~\ref{fig:concept}).
We adopted this architecture for two key reasons:
First, it manifests the critical benefit of mirroring how humans break complex tasks into subtasks, which are easier to solve. Studies in cognitive science suggest that humans use hierarchical frameworks in decision-making~\cite{botvinick2012hierarchical, frank2012mechanisms}.
Second, the machine-learning community's promising advances in implementing hierarchical architectures in various motor-control domains~\cite{huang2023inner, brohan2023can} points to its potential for oculomotor control.
This particular hierarchical architecture is composed of a high-level cognitive controller for deciding subtasks and a low-level oculomotor controller for moving the gaze.
The cognitive controller is powered by large language models (LLMs)~\cite{achiam2023gpt} for understanding the task, analyzing the information obtained, and selecting operations for collecting information from the chart.
In the oculomotor control, the approach employs deep reinforcement learning (RL)~\cite{schulman2017proximal}, training RL agents for each operation to perform detail-level gaze movements.

We evaluated the model's performance through experiments using human data. 
We compared scanpath-level similarity with baseline models' output and human scanpaths, where the baselines were the latest general scanpath prediction model~\cite{kummerer2022deepgaze}, scanpath prediction in visual question answering~\cite{chen2021predicting}, and free-viewing scanpath prediction on charts~\cite{wang2023scanpath}. The results suggest that the hierarchical gaze control model demonstrates closest similarity to human data across tasks.
We also analyze the summary statistic of the gaze behavior from model predictions, which effectively reproduce human-like gaze movement behavior.
The evaluation results highlight the potential for the model to exhibit human-like behavior in task-driven scanpaths across different tasks.

In summary, the main contribution of this work is the first computational model \name, to the best of our knowledge, for predicting task-driven scanpaths on charts. The key technical contribution of \name is its hierarchical gaze control with a cognitive controller and oculomotor controllers. This architecture enables training the model without relying on human eye movement data.
We analyzed human scanpath data from charts, to support modeling for three common analysis tasks: ``retrieve value,'' ``filter,'' and ``find extreme''.
We conducted comprehensive experiments to evaluate scanpath prediction across analytical tasks, comparing scanpath similarity and providing statistical summaries. Our model performs similarly to humans and better than the baselines in predicting task-driven scanpaths.
\rv{This study focused on analytical tasks involving statistical charts, where the scanpaths align more with the problem-solving reasoning process and are less influenced by the complexity of the visual representation. At the end of the paper, we discuss the generalizability of the modeling approach.}

\rv{
The paper is structured such that 
Section~\ref{sec:related} reviews prior research into eye tracking connected with information visualizations, analytical tasks, and scanpath prediction models 
with Section~\ref{sec:model} laying further groundwork by introducing the problem formulation and the design of the computational model for predicting task-driven scanpaths, including the hierarchical control architecture and training workflow.
We then present our evaluation of the model’s performance relative to baseline methods and human data in Section~\ref{sec:evaluation}.
Finally, Section~\ref{sec:discussion} discusses potential applications of the approach and its generalizability.
}
\section{Related Work}
\label{sec:related}

This section reviews the literature on eye tracking for information visualization, tasks in that domain, and preexisting techniques for scanpath prediction.

\subsection{Eye Tracking for Information Visualizations}

Eye tracking often serves as a proxy for visual perception in human analysis of information visualizations~\cite{shin2022scanner}.
There is a long history of applying eye tracking techniques to investigate how people perceive visualizations~\cite{huang2007using, borkin2015beyond, polatsek2018exploring, lalle2020gaze}.
For instance, \citet{huang2007using} examined the relations between eye movement events and visual components in node--link diagrams.
\citet{lalle2020gaze} analyzed the connection between gaze behavior and narrative visualizations, and
\citet{borkin2015beyond} found links between gaze behaviors over visual elements and the memorability of visualizations. For memorable visualizations, a quick look can already effectively convey the visualization's message.
Work by \citet{polatsek2018exploring}, demonstrating significant differences in gaze behaviors under three visual analysis tasks' conditions, attests well that people read completely different regions of charts when handling different tasks.
Other scholars have proposed eye-tracking-based approaches for improvements in visual analytics work such as word-sized visualizations~\cite{beck2015word} and interactive visualizations~\cite{nguyen2015interactive}.
However, while eye trackers have become cheaper and more readily available, the scale of eye tracking datasets is still limited because of the amounts of time and money needed~\cite{shin2022scanner}.
To address this limitation, researchers turned to crowdsourcing platforms as a faster, inexpensive alternative to eye tracking:
web cameras~\cite{shin2022scanner} and mouse clicks~\cite{wang2024salchartqa} have become popular ways to collect human attention data online.
These online alternatives sacrifice the quality of eye tracking data to gain quantity, leaving an open question of how to acquire both high quality and good scale of eye tracking data from information visualizations.

\subsection{Analytical Tasks in Visualizations}
Evidence exists that the task strongly influences how people design and explore visualizations~\cite{polatsek2018exploring, wu2020visact, shi2019task}.
\citet{amar2005low} identified 10 low-level analytical tasks (e.g., retrieving values and finding extremes) while another study~\cite{hibino1999task} highlighted abstract tasks such as background understanding, planning of analysis, and data exploration. 
Considering specific tasks is critical since visualizations can be created for handling any of the tasks in light of the input data, and also they can be evaluated in terms of how well certain tasks can be accomplished~\cite{schulz2013design}. 
To facilitate tasks related to visualizations, some researchers have designed visualizations for explicit displaying of data facts~\cite{srinivasan2018augmenting, shi2020calliope} -- such as showing trends or comparisons directly~\cite{shi2021autoclips, shi2024understanding}. 
Talk2data~\cite{guo2024talk2data} and Datamator~\cite{guo2023datamator} organize data facts related to specific tasks to facilitate question answering.
For this study, we adopted three analytical tasks from previous research~\cite{polatsek2018exploring} and conducted further analysis to understand how humans read charts for given tasks. 
Our analysis inspired us to develop the computational model for predicting task-driven scanpaths over charts.

\subsection{Scanpath Prediction}
In aims of predicting people's spatial and temporal viewing patterns upon exposure to certain stimuli, represented by a sequence of fixations, scholars have studied scanpaths with numerous stimulus types. Among these are natural scenes~\cite{coutrot2018scanpath, yang2020predicting}, web pages~\cite{drusch2014analysing}, graphical user interfaces~\cite{jokinen2020adaptive}, and information visualizations~\cite{wang2023scanpath}.
There is literature on scanpath prediction dedicated to sampling fixations from saliency maps\cite{brockmann2000ecology, boccignone2010gaze, kummerer2022deepgaze}. 
SaltiNet~\cite{assens2017saltinet} extended these maps to ``saliency volumes,'' from which sample scanpaths were created, while the models have drawn inspiration from cognitively plausible mechanisms, such as inhibition of return~\cite{itti1998model, sun2019visual} or foveal--peripheral saliency~\cite{wang2017scanpath, bao2020human}.
In HMM-based methods, in turn, the prediction either splits an image into several grids and regards each grid as a single state of observation~\cite{verma2019hmm} or classifies the fixations into several states~\cite{coutrot2018scanpath}.

The advent of deep learning brought new architectures into play for predicting scanpaths:
PathGAN's developers~\cite{assens2018pathgan} proposed using a generative adversarial network (GAN) for scanpath prediction, while 
Gazeformer~\cite{mondal2023gazeformer} encoded stimuli by using a natural-language model, then applied transformer-based modeling to predict visual scanpaths in a zero-shot setting.
For task-driven scanpath prediction, \citet{yang2020predicting} put inverse reinforcement learning to use to model human scanpaths during visual search, with 
\citet{chen2021predicting} likewise proposing an RL model to predict the scanpath during visual question answering.

Modeling scanpaths on information visualizations is challenging, given that viewing behaviors vary greatly across viewers~\cite{polatsek2018exploring}.
\citet{wang2023scanpath} highlighted the poor performance of prior scanpath prediction models with information visualizations, and they tackled this issue by fine-tuning a multi-duration saliency model~\cite{fosco2020predicting} for the information's graphical presentation and probabilistically sampling fixations from saliency maps.
However, their pioneering model for scanpath prediction in the visualization domain still cannot cope with task-specific scanpaths.
To fill the gap, we sought a task-driven model specifically designed to predict a human scanpath over information visualizations.
\section{\name: Modeling Task-driven Eye Movement on Charts}
\label{sec:model}

This section introduces the problem formulation and presents the computational model of eye movement control on charts in settings of analytical tasks.

\subsection{Problem Formulation}

Given a chart image $C$ and an associated analytical task $x$ stated as text, the model is expected to generate a sequence of fixation positions $\{ p_1, p_2, \dots , p_t\}$.
The objective of the output sequence is to closely match the scanpath from humans reading the chart. 
Specifically, the sequence of fixations represents the visual reasoning process, and the information in the patches of pixels fixated upon should be able to support $x$.
We consider general analytical tasks in information visualization~\cite{amar2005low}, and
select three of them used in a human eye-tracking data collection~\cite{polatsek2018exploring}: % \textit{RV}, \textit{F}, and \textit{FE} tasks
\rv{
\begin{itemize}
    \item[1)] \textit{Retrieve value (\textit{RV})}: Given a specific target, find the data value of the target (e.g., what is the value for a certain category?)
    \item[2)] \textit{Filter (\textit{F})}: Given a concrete condition, find which data point satisfies it (e.g., which category has the specific value stated?)
    \item[3)] \textit{Find extreme (\textit{FE})}: Find the data point showing an extreme value for a given attribute within the set of data (e.g., which category shows the highest/lowest value?)
\end{itemize}
}

\subsection{Modeling Overview}

\begin{figure*}[!t]
\centering
  \includegraphics[width=\textwidth]{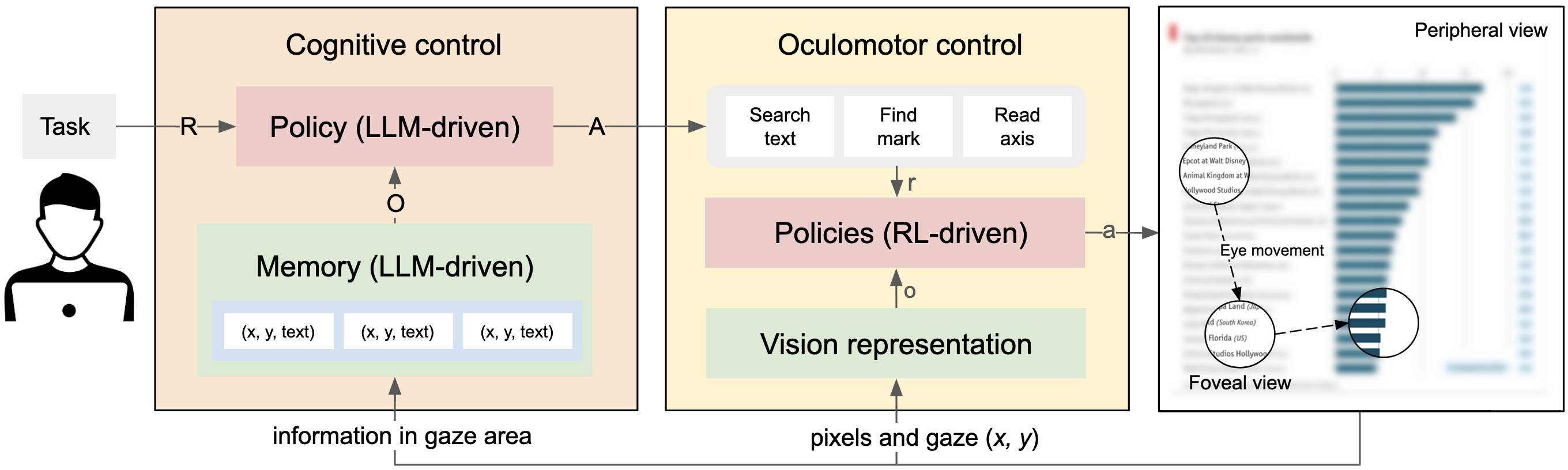}
  \caption{\textbf{An overview of the hierarchical eye-movement control architecture.} When presented with a chart and a task, a cognitive controller, powered by large language models, makes decisions on what to look at next and judges whether it is confident enough to provide an answer to the task's question. It relies on internal memory, which summarizes the information gathered from the chart through eye movements. Once cognitive control has determined the next action, the oculomotor controller is responsible for moving the gaze and observing the chart through a limited vision field. The model's objective is to accurately address the task as quickly as possible within set cognitive and physical constraints.}
  \Description{An overview of the hierarchical eye movement control architecture.}
  \label{fig:model}
\end{figure*}

Our goal was to develop the model \name to handle tasks articulated as free-form text and be able to perform gaze movement at a detailed pixel level.
We conceptualize the design of the hierarchical gaze control model in Figure~\ref{fig:model}, where the high-level (cognitive) controller is responsible for reasoning while the low-level (oculomotor) controller determines details of gaze movement. 
The idea behind this is hierarchical supervisory control~\cite{eppe2022intelligent}, which refers to a tiered control system in which the superior controller set goals for its subordinates. The actions from subordinates are integrated into an overall pattern for high-level control~\cite{pew1966acquisition}.
The concept also follows the modeling principle of computational rationality, where we assume that the controllers optimize their policy to maximize expected utility within relevant cognitive bounds~\cite{oulasvirta2022computational,chandramouli2024workflow}.
Specifically, the high-level controller handles abstract information processing, comprehension, and memory storage. 
It sets subtasks to the low-level controller, which then moves the gaze to gather information for task completion. Subsequently, the high-level controller utilizes the amassed information to answer the question.

\subsection{Cognitive Control}

The high-level controller provides cognitive control over the mental processes for a chart, control that performs reasoning in working memory~\cite{liu2010mental}. When performing vision tasks, one observes and analyzes visual information interactively~\cite{chen2020air}. Throughout this process, people analyze the information in their memory and try to gather more useful information to reduce uncertainty in solving the task.
To represent this decision problem accurately, we formulate it as a bounded optimality problem in a partially observable Markov decision process (POMDP). Instead of having access to a full state ($\mathcal{S}$) with pixels of the chart associated with the given task, the POMDP expresses a subset of ($\mathcal{S}$) as the observation of the model:
\begin{itemize}
    \item Observation $O$ refers to the information in memory that is captured from eye movements over the chart.
    \item Action $A$ includes subtasks that the model gives to oculomotor control for performing eye movements.
    \item Reward $R$ is the correctness of the answer for the task from the chart question answering.
\end{itemize}
To solve this POMDP, our model uses LLMs for the policy. The rationale behind this choice is that LLMs are well suited to processing higher-level information, as they have been pre-trained on human text data encompassing a wealth of logic related to planning, reasoning, and interaction~\cite{huang2022language, vemprala2024chatgpt, li2023interactive}.
Although LLMs are limited in their ability to control low-level motor functions in a precise manner~\cite{dalal2024psl}, they are proficient at planning and reasoning, with LLaMA~\cite{touvron2023llama} and GPT~\cite{achiam2023gpt} showing impressive language interpretation and reasoning capabilities.
Also, recent work has shown that utilizing LLMs in the high-level controllers in hierarchical architecture can produce promising results~\cite{huang2022language, brohan2023can, liang2023code}.
For our setting, we used GPT-4o~\cite{achiam2023gpt} for the policy, which takes the information accumulated in the memory as the observation and sets subtasks to guide eye movements in order to obtain information needed for solving the task efficiently.

We consider two human limitations when constructing the model's observation: a limited field of vision~\cite{duchowski2018gaze} and memory capacity~\cite{loftus2019human}. 
The model gets information from the gaze position purely by mimicking the human vision system. 
An optical character recognition technique~\cite{singh2010optical} is used to extract text from the pixels of the chart, and the text in the gaze area, with the position, is passed to the memory.
As a result, the observation consists of image patches (in a limited number) from the full set of chart pixels. The reliability of items in memory is determined by their visit history~\cite{li2023modeling}, with overall memory capacity being restricted too. When new information is added to the memory, a previously added item is removed on the basis of a forgetting probability. The probability of forgetting an item in the memory is calculated by means of the formula $\text{Softmax}(\rho \cdot (t-t_i)) $, where $t$ is the current fixation index, $t_i$ is the index of the $i$th item in the memory, and $\rho$ is the weight parameter (set to 0.1 here). The observation is designed as a prompt that summarizes the memory in line with the memory model and explains the model's goal. 

Given the summary of the memory information, the LLM policy selects predefined operations for task solving~\cite{brohan2023can, liang2023code}. The operations here are based on a sequence of cognitive stages for charts~\cite{goldberg2011eye} -- 1) \textit{search for text label}: visually searching for a text label or value label related to the task, 2) \textit{find associated mark}: visually searching for a graphical mark of the data point when given a reference label, 3) \textit{read associated value}: visually searching to read the given mark's associated value or textual label.
All these actions are allowed to be reused in the process, which enables the model to revisit previous positions for confirmation of the information.
Ultimately, if the information in the memory is sufficient to address the task, the gaze movement can stop and an answer can be given. Operations other than answering the question will be performed by the oculomotor controller for detailed gaze movement. 

\begin{figure*}[!t]
\centering
  \includegraphics[width=\textwidth]{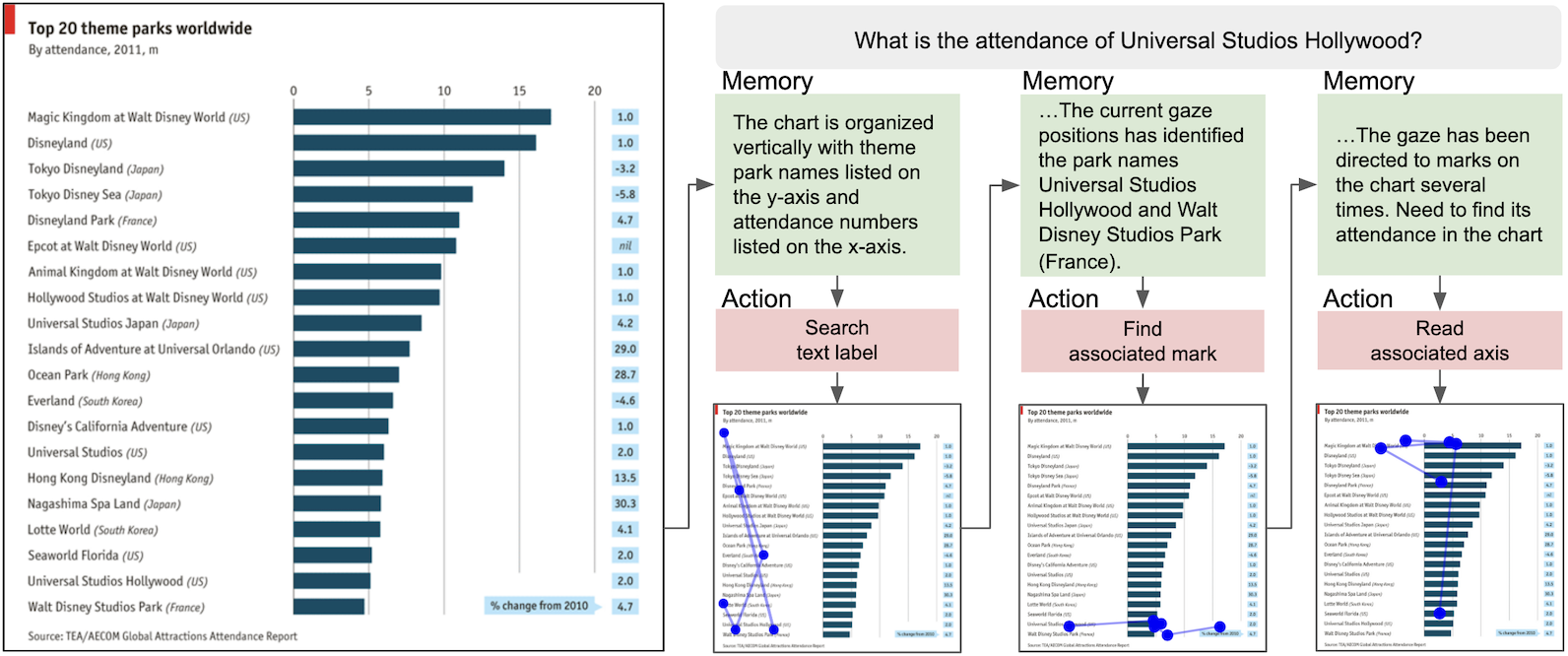}
  \caption{The figure gives examples of how the internal memory helps the cognitive controller to remember what has been read and then select actions for detailed gaze movement. A green box indicates the information held in memory, a red box represents the action selected by cognitive control, and the blue lines in the images reflect the eye movement scanpaths.}
  \label{fig:memory}
\end{figure*}

The examples in Figure~\ref{fig:memory} demonstrate how utilizing memory information and predefined operations aids in scanpath prediction. Model memory uses the summarization capability of LLMs to convert the text and positions gathered to a paragraph as the observation (as shown in the green boxes). The LLM policy then makes decisions and issues subtasks as actions (in red boxes) for the oculomotor control, which performs pixel-level gaze movements.

\subsection{Oculomotor Control}

The oculomotor controller acts as the interface between the cognitive controller and the actual chart-pixel images. Its main function is to control the movement of the gaze over the pixels in order to gather information related to the task at hand.
Generating oculomotor behavior at pixel level is another sequential decision-making problem that can be formulated as a POMDP:
\begin{itemize}
    \item Observation $o$ comprises vision information obtained from the external environment, which is jointly represented by the human vision system and visual short-term memory (VSTM).
    \item Action $a$  involves specifying the coordinates $(x, y)$ of a particular position to move to.
    \item Reward $r$ is designed to encourage the gaze to reach the target with less cost. It takes into account the number of target hits as well as the cost associated with the distance of the gaze movement.
\end{itemize}

Our modeling of a chart reader's observation follows an idea similar to that in visual search~\cite{yang2020predicting}. Utilizing a representation for accumulating information through fixations, this employs four components: 
1) The foveal and peripheral view come from the human vision system, which receives high-resolution visual input only from the region of the image around the fixation location. It includes two pixel-based modules to read the chart: foveal and peripheral vision~\cite{duchowski2018gaze}). 
2) Visual saliency provides a bottom-up signal to a chart reader for the given task. The saliency of the chart affects gaze behavior. We use a task-driven saliency model to represent this feature ~\cite{wang2024salchartqa}.
3)  Visit history represents VSTM, which stores visual information for a few seconds, thereby allowing its use in ongoing cognitive tasks~\cite{alvarez2004capacity}. We represent this history through a matrix where each point is marked as visited or not.
4) A goal-related reference position serves as the initial starting point of gaze movement. For example, the reader might begin at the position of a text label for locating the associated graphical mark, where the position of the text label serves as the reference for the sub-goal. 
\rv{We use a one-hot matrix to represent the reference, in which all cell values are 0 apart from the single 1 that identifies the target.}
All these components are encoded together via the deep convolutional neural network, followed by a fully connected network.

We train reinforcement learning policies to solve the POMDP for the oculomotor control, because it has been proven to effectively address decision-making challenges in prediction of details of gaze movement~\cite{yang2020predicting, jiang2024eyeformer, shi2024crtypist, bai2024heads}.
In our detail-level implementation, we resize the input chart images to be $320 \times 320$ and discretize the fixation position into a $20 \times 20$ map. Consequently, each fixation becomes a $16 \times 16$ image patch, and the gaze position is randomly sampled from within that patch. In this setup, the maximum approximation error resulting from this discretization process is less than one degree of the visual angle~\cite{yang2020predicting}.
\rv{
Ultimately, both the scanpath and the image will be converted back to the original chart size from $320 \times 320$ pixels.
}

\subsection{\rv{Workflow}}
\label{sec:workflow}

\begin{figure*}[!h]
\centering
  \includegraphics[width=\textwidth]{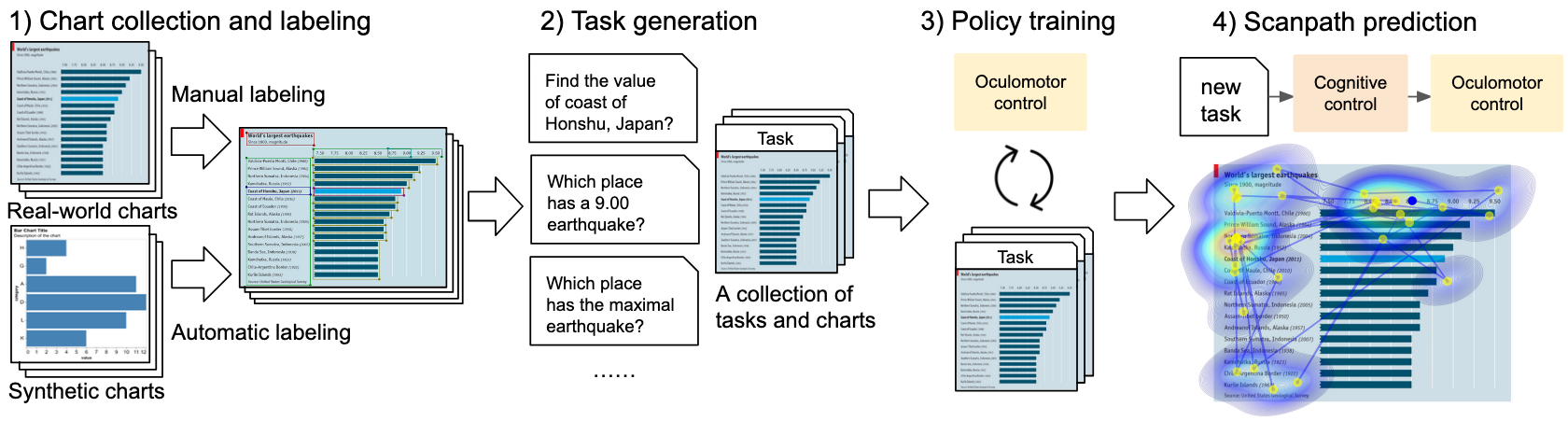}
  \caption{\rv{An overview of the training workflow: 1) chart collection and labeling, wherein diverse real-world and synthetic charts are gathered, involving manual and automatic annotation of AOIs; 2) task generation, utilizing a rule-based approach to create tasks based on labeled charts to construct a data collection for training; 3) policy training, in which policy models are trained via RL from chart images with tasks; and 4) scanpath prediction, wherein pre-trained LLMs and RL policies are coordinated hierarchically to predict task-driven gaze movements over charts.}}
  \label{fig:pipeline}
  % \vspace{-10mm}
\end{figure*}

\rv{
Our implementation of \name is trained and tested on a collection of tasks and charts. There are four steps, illustrated in Figure~\ref{fig:pipeline}.
In Step 1, real-world charts are manually collected and labeled for areas of interest (AOIs), while synthetic charts are automatically generated and labeled in a manner powered by Vega-Lite~\cite{satyanarayan2016vega}. The inclusion of synthetic charts helps increase the diversity of the chart collection and addresses the challenge of obtaining numerous annotated charts.
In Step 2, tasks are automatically generated in line with specific rules for the \textit{RV}, \textit{F}, and \textit{FE} tasks. These tasks and labeled charts constitute a data collection for the training environment.t
With Step 3, the policies for oculomotor control are trained through reinforcement learning (using proximal policy pptimization, PPO~\cite{schulman2017proximal}) to optimize gaze movements, enabling the system to reach task-relevant positions as quickly as possible while adhering to vision constraints. Importantly, no eye tracking data are required for PPO training.
In the last phase, prediction, the hierarchical architecture combines pre-trained LLMs (GPT-4o) for cognitive control with RL policies for oculomotor control to generate the scanpath prediction.
}

\section{Experiments}
\label{sec:evaluation}

This section presents the experiments conducted for evaluating and comparing scanpath prediction models. We evaluated \name specifically in terms of scanpath similarity and statistical summaries of eye movement behavior. The results from evaluation of our model in comparison to the baselines are summarized in Table~\ref{tab:benchmark}.

\subsection{Data and Metrics}

We evaluated \name by using 12 distinct analytical tasks with horizontal bar charts from a task-driven scanpath dataset~\cite{polatsek2018exploring}. 
Each task has at least 14 human scanpaths (\textit{M} = 15.25, \textit{SD} = 0.60), for a total of 183 human scanpaths. 
\rv{
This set of ground-truth human data was collected by means of Tobii X2-60 eye trackers at 60~Hz while participants were engaged in these analytical tasks on 24.1-inch monitors at a resolution of 1920 × 1080 pixels.
}
To evaluate model performance, we compared the generated scanpaths against human ground truth, with the aim of ascertaining whether the models can produce human-like scanpaths and replicate natural gaze movement patterns. 
We should stress that, to ensure unbiased evaluation, none of the human eye movement data informed our training of the models, only their assessment.
Furthermore, except for scanpath prediction that relies on human-generated data as the training set, no such eye movement data were involved in training for our approach.

Following recent practice in scanpath prediction~\cite{wang2023scanpath, sui2023scandmm}, we evaluated \name by using three established scanpath-based metrics: dynamic time warping (DTW), Levenshtein distance (LEV), and Sequence Score.
DTW computes the optimal alignment between two scanpaths, where lower values indicate better correspondence. For this paper, DTW was calculated in two-dimensional position coordinates.
Both LEV and Sequence Score represent the semantic order of scanpaths as sequences of letters by mapping each fixation to a unique letter, then measuring the string-editing distance between the sequences~\cite{needleman1970general}. 
In LEV, letters are defined by the grid regions on which fixations land. 
For Sequence Score~\cite{yang2020predicting, wang2023scanpath}, letters are based on areas of interest (AOIs) such as the title and legend.
Sequence Score values are normalized between 0 and 1, with higher scores reflecting better alignment.
For DTW, LEV, and Sequence Score, we report the \textit{mean} and \textit{best} evaluation scores (see Table 2). For each method, the number of scanpaths equaled that of human scanpaths. The \textit{mean} scores are the averages across all human--predicted scanpath pairs, while the \textit{best} ones represent the maximum of all pairs for each prediction~\cite{chen2021predicting, wang2023scanpath}.

\definecolor{best}{rgb}{0.63137, 0.85098, 0.60784}
\definecolor{good}{rgb}{0.89804, 0.96078, 0.87843}

% \textbf{Andreas: I suggest to move the "Ours" column forward, before "Human". makes comparing numbers between Ours and Human easier. And doesn't "hide" the most important results at the end; add thousand delimiter}

\begin{table*}[htbp]
\large
\centering
\caption{A quantitative benchmark of the task-driven human scanpaths on bar charts. The best results are shown in dark green. All results within 1 standard deviation from human are in light green.}
\begin{tabular}{cl p{15mm} p{15mm}p{15mm}p{15mm}p{15mm}}
\toprule
Task & Metric & Human & \name & VQA~\cite{chen2021predicting} & UMSS~\cite{wang2023scanpath} & DG iii~\cite{kummerer2022deepgaze} \\
\midrule 
\multirow{15}*{\makecell[l]{Retrieve value}} 
& Sequence Score~$\uparrow$ (\textit{mean}) & 0.486 & {\cellcolor{best}} \textbf{0.413} & 0.127 & 0.246 & 0.345 \\ % & {\cellcolor{good}} \textbf{0.380}\\
& Sequence Score~$\uparrow$ (\textit{best}) & 0.638 & {\cellcolor{best}} \textbf{0.472} & 0.151 & 0.312 & 0.404 \\ %{\cellcolor{good}} \textbf{0.484}\\
& LEV~$\downarrow$ (\textit{mean}) & 154.7 & {\cellcolor{best}} \textbf{151.8} & 164.6 & 157.4 & 189.7 \\ % 188.4\\
& LEV~$\downarrow$ (\textit{best}) & 121.5 & {\cellcolor{best}} \textbf{141.7} & 162.5 & 150.9 & 181.4 \\ %153.0\\
& DTW~$\downarrow$ (\textit{mean}) & 26,018 & 28,172 & 36,703 & {\cellcolor{best}} \textbf{26,305} & 37,160 \\
& DTW~$\downarrow$ (\textit{best}) & 17,692 & 23,541 & 33,121 & {\cellcolor{best}} \textbf{21,081} & 30,574 \\ %25081\\
~ & Number of fixations        & 88.6\,(57.0) & {\cellcolor{best}} \textbf{48.4}\,(4.9) & 8.5\,(0.5) & 22.2\,(3.7) & -\\
~ & Task AOIs ratio (\%)       & 10.4\,(7.8) & {\cellcolor{good}} \textbf{4.1}\,(4.3) & 1.2\,(3.9) & {\cellcolor{good}} \textbf{3.8}\,(5.8) & {\cellcolor{best}} \textbf{5.8}\,(8.2)\\
~ & Fixation-on-title ratio (\%)   & 10.5\,(9.3) & {\cellcolor{good}} \textbf{5.3}\,(3.3) & 2.8\,(5.7) & {\cellcolor{best}} \textbf{13.7}\,(8.7) & 33.5\,(13.4)\\
~ & Fixation-on-mark ratio (\%)    & 31.5\,(13.9) & {\cellcolor{best}} \textbf{38.6}\,(21.3) & 64.0\,(24.3) & 51.1\,(12.8) & 15.3\,(10.3)\\
~ & Fixation-on-axis ratio (\%)    & 47.5\,(19.0) & {\cellcolor{best}} \textbf{43.9}\,(17.2) & 23.8\,(20.8) & 21.5\,(14.7) & {\cellcolor{good}} \textbf{28.6}\,(10.5)\\
~ & Fixation transitions & 20.1\,(12.4) & {\cellcolor{best}} \textbf{16.7}\,(7.0) & 3.4\,(1.7) & {\cellcolor{good}} \textbf{10.6}\,(3.3) & 33.9\,(9.2)\\
~ & Revisit frequency title    & 2.6\,(2.4) & {\cellcolor{good}} \textbf{2.0}\,(1.3) & {\cellcolor{good}} \textbf{0.2}\,(0.5) & {\cellcolor{best}} \textbf{2.1}\,(1.2) & 10.1\,(2.6)\\
~ & Revisit frequency mark     & 7.9\,(5.1) & {\cellcolor{good}} \textbf{6.8}\,(3.4) & 1.8\,(0.8) & {\cellcolor{good}} \textbf{4.3}\,(1.5) & {\cellcolor{best}} \textbf{8.6}\,(4.4)\\
~ & Revisit frequency axis     & 7.9\,(4.4) & {\cellcolor{best}} \textbf{7.8}\,(3.4) & 1.3\,(1.0) & 3.0\,(1.9) & {\cellcolor{good}} \textbf{12.4}\,(3.2)\\
\midrule 
\multirow{15}*{\makecell[l]{Filter}} 
& Sequence Score~$\uparrow$ (\textit{mean}) & 0.452 & {\cellcolor{best}} \textbf{0.379} & 0.149 & 0.271 & 0.321 \\ %{\cellcolor{good}} \textbf{0.357} \\
& Sequence Score~$\uparrow$ (\textit{best}) & 0.655 & {\cellcolor{best}} \textbf{0.460} & 0.175 & 0.340 & 0.382 \\ % {\cellcolor{good}} \textbf{0.501}\\
& LEV~$\downarrow$ (\textit{mean}) & 165.8 & {\cellcolor{best}} \textbf{155.3} & 167.6 & 161.0 & 201.5 \\ %187.6\\
& LEV~$\downarrow$ (\textit{best}) & 114.5 & {\cellcolor{best}} \textbf{146.4} & 164.7 & 153.5 & 192.6 \\ %{\cellcolor{good}} \textbf{139.3}\\
& DTW~$\downarrow$ (\textit{mean}) & 23,732 & 24,617 & 29,141 & {\cellcolor{best}} \textbf{22,817} & 38,139 \\ %32243\\
& DTW~$\downarrow$ (\textit{best}) & 15,238 & 19,879 & 25,233 & {\cellcolor{best}} \textbf{17,725} & 30,981 \\ %23259\\
& Number of fixations & 89.1\,(68.7) & {\cellcolor{best}} \textbf{48.7}\,(3.3) & 8.5\,(0.5) & {\cellcolor{good}} \textbf{22.2}\,(3.7) & -\\
~ & Task AOIs ratio (\%) & 19.1\,(14.4) & {\cellcolor{good}} \textbf{10.6}\,(11.2) & {\cellcolor{best}} \textbf{13.9}\,(17.4) & 3.8\,(5.8) & {\cellcolor{good}} \textbf{5.8}\,(8.2)\\
~ & Fixation-on-title ratio (\%) & 5.5\,(5.7) & {\cellcolor{best}} \textbf{4.7}\,(2.3) & {\cellcolor{good}} \textbf{2.0}\,(5.0) & 13.7\,(8.7) & 33.5\,(13.4)\\
~ & Fixation-on-mark ratio (\%) & 41.9\,(18.4) & {\cellcolor{best}}
\textbf{49.1}\,(21.8) & 68.8\,(23.5) & {\cellcolor{good}} \textbf{51.1}\,(12.8) & 15.3\,(10.3)\\
~ & Fixation-on-axis ratio (\%) & 43.3\,(20.9) & {\cellcolor{best}} \textbf{36.7}\,(19.4) & 20.6\,(19.4) & 21.5\,(14.7) & {\cellcolor{good}} \textbf{28.6}\,(10.5)\\
~ & Fixation transitions & 18.4\,(13.3) & {\cellcolor{best}} \textbf{15.3}\,(5.8) & 3.2\,(1.7) & {\cellcolor{good}} \textbf{10.6}\,(3.3) & 33.9\,(9.2)\\
~ & Revisit frequency title & 1.4\,(1.7) & {\cellcolor{best}} \textbf{2.0}\,(1.1) & {\cellcolor{good}} \textbf{0.2}\,(0.4) & {\cellcolor{good}} \textbf{2.1}\,(1.2) & 10.1\,(2.6)\\
~ & Revisit frequency mark & 7.7\,(5.6) & {\cellcolor{good}} \textbf{6.0}\,(2.9) & 1.8\,(0.8) & {\cellcolor{good}} \textbf{4.3}\,(1.5) & {\cellcolor{best}} \textbf{8.6}\,(4.4)\\
~ & Revisit frequency axis & 8.0\,(5.5) & {\cellcolor{best}} \textbf{7.0}\,(2.5) & 1.2\,(1.1) & {\cellcolor{good}} \textbf{3.0}\,(1.9) & {\cellcolor{good}} \textbf{12.4}\,(3.2)\\
\midrule
\multirow{15}*{\makecell[l]{Find extreme}}
& Sequence Score~$\uparrow$ (\textit{mean}) & 0.454 & {\cellcolor{best}} \textbf{0.378} & 0.126 & 0.253 & 0.362 \\ %{\cellcolor{good}} \textbf{0.364}\\
& Sequence Score~$\uparrow$ (\textit{best}) & 0.627 & {\cellcolor{best}} \textbf{0.457} & 0.149 & 0.320 & 0.428\\ %{\cellcolor{good}} \textbf{0.496}\\
& LEV~$\downarrow$ (\textit{mean}) & 167.3 & {\cellcolor{best}} \textbf{151.4} & 168.3 & 160.5 & 188.9 \\ %176.4\\
& LEV~$\downarrow$ (\textit{best}) & 123.9 & {\cellcolor{best}} \textbf{143.9} & 168.3 & 155.6 & 180.9 \\ % {\cellcolor{good}} \textbf{151.2}\\
& DTW~$\downarrow$ (\textit{mean}) & 27,701 & 26,677 & 34,287 & {\cellcolor{best}} \textbf{25,398} & 36,878 \\ %32728\\
& DTW~$\downarrow$ (\textit{best}) & 18,626 & 22,537 & 31,379 & {\cellcolor{best}} \textbf{20,631} & 30,400 \\ % 25960\\
~ & Number of fixations        & 91.9\,(64.9) & {\cellcolor{best}} \textbf{46.2}\,(6.8) & 8.5\,(0.5) & 22.2\,(3.7) & -\\
~ & Task AOIs ratio (\%)       & 1.7\,(3.4) & {\cellcolor{best}} \textbf{0.2}\,(0.7) & 0\,(0) & {\cellcolor{good}} \textbf{3.8}\,(5.8) & 5.8\,(8.2)\\
~ & Fixation-on-title ratio (\%) & 12.0\,(8.6) & {\cellcolor{good}} \textbf{5.4}\,(3.2) & 2.4\,(5.5) & {\cellcolor{best}} \textbf{13.7}\,(8.7) & 33.5\,(13.4)\\
~ & Fixation-on-mark ratio (\%)  & 34.4\,(19.4) & {\cellcolor{best}} \textbf{41.6}\,(15.6) & 64.7\,(25.6) & {\cellcolor{good}} \textbf{51.1}\,(12.8) & {\cellcolor{good}} \textbf{15.3}\,(10.3)\\
~ & Fixation-on-axis ratio (\%)  & 39.7\,(22.8) & {\cellcolor{best}} \textbf{40.7}\,(11.3) & {\cellcolor{good}} \textbf{23.7}\,(22.9) & {\cellcolor{good}} \textbf{21.5}\,(14.7) & {\cellcolor{good}} \textbf{28.6}\,(10.5)\\
~ & Fixation transitions        & 22.6\,(16.8) & {\cellcolor{best}} \textbf{16.0}\,(4.9) & 3.0\,(1.6) & {\cellcolor{good}} \textbf{10.6}\,(3.3) & {\cellcolor{good}} \textbf{33.9}\,(9.2)\\
~ & Revisit frequency title     & 3.2\,(3.3) & {\cellcolor{good}} \textbf{2.0}\,(1.1) & {\cellcolor{good}} \textbf{0.2}\,(0.5) &{\cellcolor{best}} \textbf{2.1}\,(1.2) & 10.1\,(2.6)\\
~ & Revisit frequency mark      & 8.1\,(5.6) & {\cellcolor{good}} \textbf{6.1}\,(2.4) & 1.7\,(0.8) & {\cellcolor{good}} \textbf{4.3}\,(1.5) & {\cellcolor{best}} \textbf{8.6}\,(4.4)\\
~ & Revisit frequency axis      & 8.9\,(5.7) & {\cellcolor{best}} \textbf{7.7}\,(2.3) & 0.8\,(1.1) & 3.0\,(1.9) & {\cellcolor{good}} \textbf{12.4}\,(3.2)\\
\bottomrule
\end{tabular}
\vspace{5mm}
\label{tab:benchmark}
\end{table*}

\rv{
We looked beyond scanpath metrics, introducing more detailed measurements inspired by \citet{goldberg2010comparing} to show a statistical summary of task-driven scanpaths over charts. % The analysis of human data can be summarized thus:
\begin{itemize}
    \item \textit{Number of fixations}: The length of a scanpath can be measured as the count of gaze fixations (between motions, or saccades). According to the human data, the number of fixations in task-driven scanpaths over charts (89.8 on average) is much larger than the number in free-viewing tasks (37.4 on average). This reflects the difficulty of analytical tasks relative to free viewing of charts.
    \item \textit{Fixation on task-dependent AOI ratio}: Task-dependent AOIs are regions that are relevant to the task, such as value labels, text labels, and data points~\cite{polatsek2018exploring}. People's focus on these areas indicates how they are processing the task. Inspired by the Hit Any AOI Rate metric~\cite{wang22_etvis}, this measurement provides a summary of the overall visual attention to task-related regions. Although the scanpath is task-driven, we ascertained that most of the eye movement does not occur in task-dependent regions: fewer than 20\% of fixations fell in task-dependent AOIs. This suggests that people might devote more time to gathering information or confirming it.
    \item \textit{Percentage of fixations within each area}: We considered the percentage of time devoted to looking at distinct parts of a chart -- namely, the key areas of charts: the title, the marks (such as bars or data points), and the axes. This assists in summarizing where people are focusing their visual attention. The percentages are calculated by dividing the number of fixations in a specific area by the total number of fixations. Humans direct most of their fixations to the axes, then the region of graphical marks. This might be because the three analysis tasks probed are strongly related to values, not other visual features.
    \item \textit{Fixation transitions}: We also used a metric capturing the average number of times the eyes move from one area to another during a task. It helps us understand how often the eyes' fixations shift between distinct areas. Frequent fixation transitions may point to room for improvement in the design of the chart, such as bringing related elements closer together. From human data, we identified a high number of fixation transitions (about 20 per task). We found that, on average, about four consecutive fixations follow each fixation transition.
    \item \textit{Revisit frequencies}: The average number of fixations returning to a previously visited area during a task proved similarly revealing. Human data exhibited high revisit rates. Spatially, users revisit marks and also axes eight times, on average. This frequent double-checking of data information in the chart for the answer may be due to forgetting the information.
\end{itemize}
These metrics help us evaluate whether the model's predictions can accurately capture general human patterns followed with charts for particular tasks. We strove for a system in which the predicted scanpath closely matches human ground-truth performance, ideally being within one standard deviation of the mean value.
}

\subsection{Comparison Methods}

Given the lack of existing methods for predicting task-driven scanpaths on information visualizations, we compare \name against human ground truth with three closely related baselines:

\begin{itemize}
    \item \textit{Human}~\cite{polatsek2018exploring}. With the scanpath metrics, we conducted leave-one-out cross-validation among the human scanpaths. For each viewing condition, every human scanpath was compared with all other human scanpaths for similarity. Human scanpaths were compared with themselves for the \textit{mean} scores but not for contributions to the \textit{best} scores. In applying the statistical metrics, we treated all the human data as the ground truth and gauged all modeling methods by their closeness to this ground truth.
    \item \textit{VQA scanpaths}~\cite{chen2021predicting}. VQA is a deep reinforcement learning model that predicts human visual scanpaths in the context of images with visual question answering. The paper reporting on it demonstrates its strong generalizability across various tasks and datasets, indicating optionality as an approach for predicting task-driven scanpaths over charts. % It allows for comparing stimuli effects on scanpaths.
    \item \textit{UMSS}~\cite{wang2023scanpath}. UMSS represents the state-of-the-art scanpath prediction model for visualizations, making it the most relevant work in this area. However, it is designed to predict scanpaths in a free-viewing context for information visualizations, rather than consider specific tasks. Its inclusion allows for comparison between scanpath prediction with and without task-linked factors.
    \item \textit{DeepGaze iii}~\cite{kummerer2022deepgaze}. DeepGaze iii is a deep-learning-based model that integrates image data with information about previous fixations to forecast free-viewing scanpaths over static images. Trained on large sets of eye tracking data from natural images, it serves as a baseline for evaluating the effect both of stimuli and of tasks on scanpaths.
\end{itemize}

\subsection{Results}

\begin{figure*}[!t]
\centering
  \includegraphics[width=0.95\textwidth]{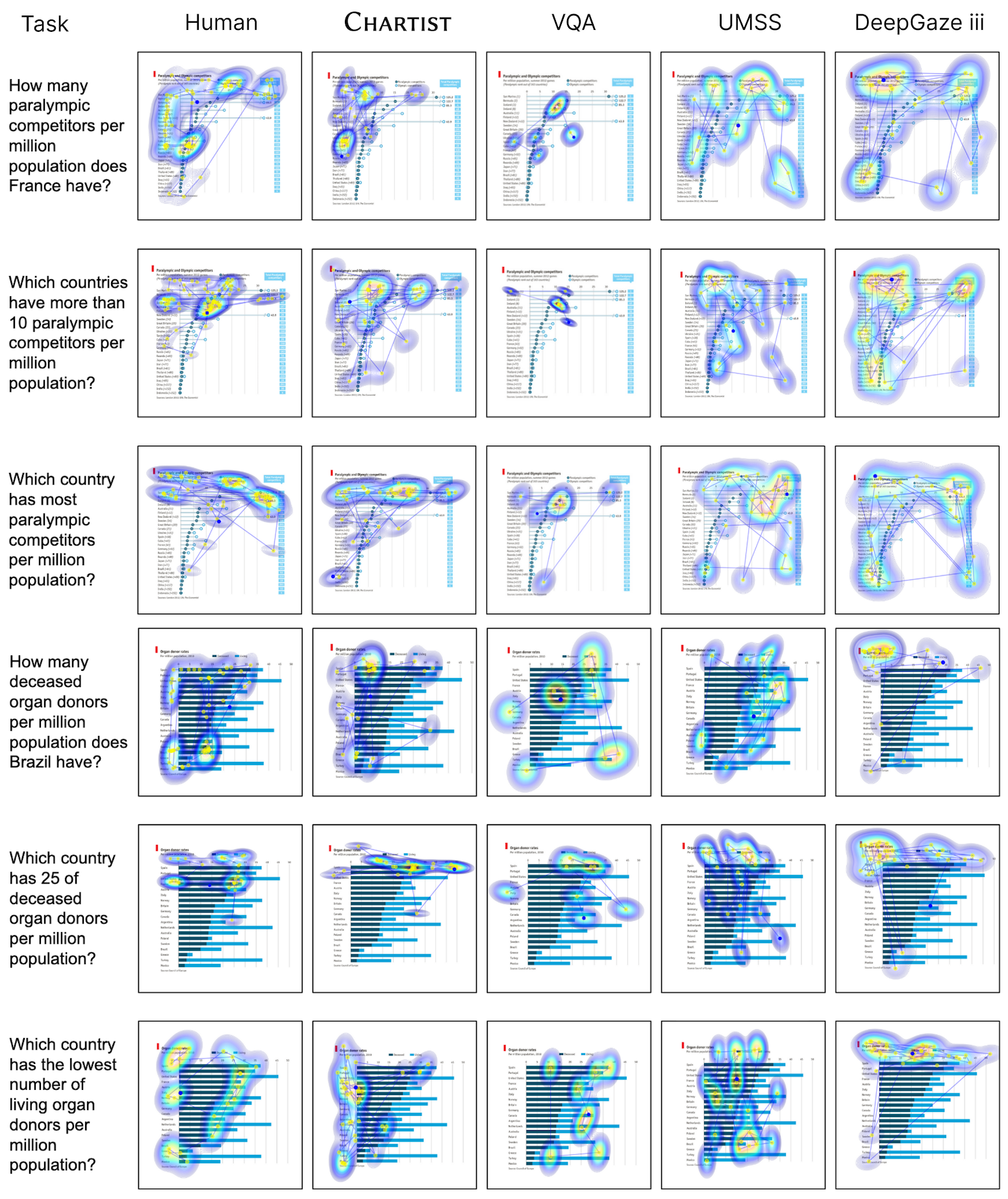}
  \caption{Qualitative comparison: for three tasks, an illustration of \name's predictions relative to three baselines -- VQA scanpaths~\cite{chen2021predicting}, UMSS~\cite{wang2023scanpath}, and DeepGaze iii~\cite{kummerer2022deepgaze}. \name is able to capture human scanpath patterns displayed during analytical tasks.}
  \label{fig:examples}
\end{figure*}

% \paragraph{Scanpath similarity}
\paragraph{\name demonstrates high similarity in scanpaths}

The first six rows for each task type in \autoref{tab:benchmark} present the results from our three scanpath similarity metrics.
\name achieved the highest performance by the Sequence Score and LEV metrics, and it ranked second for DTW, with scores closely approaching the maximum and also close to human ground truth. 
Specifically, \name closely approximates the latter Sequence Score in terms of mean performance, achieving a score of 0.413, relative to 0.486.
The UMSS method, while securing first place for DTW, ranked second for LEV and third for Sequence Score.
\name outperforms human ground truth in LEV \textit{mean} (151.8 vs. 154.7). This means that the predictions from \name deviate less from the ``average human scanpath.''

The results from the scanpath metrics show that \name performed better than the baselines by the Sequence Score and LEV metrics, which are based on regions, but not the DTW metric, which is based on pixel-wise distances. This suggests that \name is more similar to human data when one factors in the semantic order of fixation positions in meaningful portions of charts. However, it may not fully match human data for pixel-level similarity.
This result is consistent with the discussion in the literature~\cite{wang2023scanpath}, which has concluded that metrics based on pixel-wise distances between scanpaths might not wholly capture the quality of human scanpaths. Therefore, we must conduct further analysis of the statistical summary of eye movement behaviors.

% \paragraph{Statistical summary of eye movement behaviors}
\paragraph{\name aligns more closely with human statistical patterns}

The last nine rows for each task type in \autoref{tab:benchmark} provide the mean and standard deviation for each eye movement behavior. 
Because UMSS and DeepGaze iii are not task-driven, our analysis used the same predicted scanpaths across all tasks.
\name achieves strong alignment with human data, with all 27 of its values for the eye movement behavior metrics falling within one standard deviation of the human mean and with 18 of them being the closest to the human data's mean. In comparison, 18 of UMSS's 27 values lie within one standard deviation, and five of them are the closest to the mean. The corresponding figures for DeepGaze iii are 13 out of 27 and 4, respectively, while VQA yielded only six values within the range and only one of the 27 was the closest to the human mean.

Examining the detailed metrics across tasks reveals that humans show significantly variable task-dependent AOI ratios. They devote the majority of their fixations to task AOIs when performing the \textit{F} task (19.1\%). That is followed by the \textit{RV} task (10.4\%), with the \textit{FE} task having the lowest percentage (1.7\%). This distribution makes sense: the first two tasks require individuals to focus on a specific data label, while \textit{FE} can be completed by directly observing the general shape of the graph. \name is the only model that successfully replicates this phenomenon by reproducing the human order of task-dependent AOI ratios: {FE} (10.6\%), then {RV} (4.1\%), and finally {FE} (0.2\%). 
As for per-region fixation ratios, humans direct the most fixations to axes, followed by marks, across all three tasks. \name successfully reproduces this phenomenon in the case of the \textit{RV} task. For the \textit{F} and \textit{FE} tasks, \name shows similar distributions. In contrast, the VQA and UMSS baselines consistently allocate over 50\% of fixations to the marks, and DeepGaze iii allocates most fixations to the title. 
In the realm of revisits, \name and DeepGaze iii align with human data, revisiting the axes most frequently, while the other two models revisit the marks most often. 
In summary, our analysis demonstrates that \name exhibits the pattern most similar to human data.

\paragraph{Qualitative analysis}

Figure~\ref{fig:examples} showcases predicted scanpaths from \name and the three baseline models across six tasks, with fixation density maps overlaid. In all cases, our model's predictions are closer to the human data than the baseline models'. \name and VQA scanpaths both are task-driven, unlike UMSS and DeepGaze iii's, which cannot predict scanpaths solely from images. Here are the main observations from Figure~\ref{fig:examples}:
\begin{itemize}
    \item DeepGaze iii predicts scanpaths from the bottom up on the basis of the saliency of a natural image. That results in a high number of fixations on the title, consistent with producing the highest fixation-on-title ratios. 
    \item Although UMSS is not task-driven, its predicted scanpath shape closely mirrors human data. This indicates that, even in task-driven scenarios, bottom-up mechanisms exert a significant influence. 
    \item Nevertheless, \name, as does VQA, captures human gaze patterns more effectively across tasks than these free-viewing models. Importantly, \name outperforms the VQA scanpath model, for VQA often predicts fixating on irrelevant areas, in the absence of specific knowledge of visualization structures.
\end{itemize}

The examples in Figure~\ref{fig:teaser} demonstrate how \name's predictions compared to human data for the three tasks. The chart read displays a list of top-ranked theme parks worldwide with their corresponding attendance numbers for 2011. 
When given the \textit{RV} task of answering ``what is the attendance level of Universal Studios Hollywood?'' both the human user and the model focus on the text labels to find the theme park in the chart and the relevant positions for the mark and on the value axis.
For the \textit{F} task, the human user and the model both frequently look at the value axis.
Regarding the \textit{FE} task, both human and model focus on the top of the mark and also fixate on the text label associated with that mark. 
We noted that human eye movements are also drawn to other text labels, such as annotations and textual descriptions, while \name remains task-focused without getting distracted by unrelated information.
\section{Discussion}
\label{sec:discussion}

\rv{
While the results show that \name is able to simulate human-like eye movements when performing analytical tasks, there is a need to expand on our discussion of the model's practical implications, the generalizability of the modeling approach, and the limitations and potential for supporting sophisticated chart-based question answering.
}

\subsection{\rv{Applications}}

\rv{
\paragraph{Visualization design evaluation}
\name can assist in evaluation of chart design.
With well-controlled experiment conditions, eye tracking data afford valuable insight into chart designs, especially relative to alternative designs.
For example, \citet{goldberg2011eye} showcased eye tracking's value in comparing line and radial graphs for reading of values, by allowing researchers to understand the viewing order of AOIs and the task completion time.
\name holds potential to replace human input to evaluation based on eye tracking.
With the simulated scanpaths from \name, chart designers can obtain quick and cost-effective feedback that yields the benefits from eye tracking without requiring an expensive empirical study.
}
\rv{
\paragraph{Visualization design optimization}
Beyond evaluation, another potential usage application of \name is to help optimize visualization design~\cite{shin2023perceptual}. 
Like other fields of design, visualization design requires user feedback for continual iteration. When visualization designers create charts for specific tasks, they may wonder if the design is suitable for delivery.
With the predicted scanpaths from the model, they can easily access quick and affordable feedback before deeming a candidate design ready for expensive evaluation in a user study.
Predictive models could offer feedback to designers or even provide optimization goals in automated visualization design frameworks.
The ultimate goal is grounding for recommendations for visualizations that support specific tasks~\cite{albers2014task} and even automation of visualization design in real time.
Today's human-in-the-loop design optimization paradigm~\cite{kadner2021adaptifont} could shift to a user-agent-in-the-loop approach, wherein a computational agent that simulates human feedback enables scalable and efficient design evaluation.
}

\rv{
\paragraph{Explainable AI in chart question answering}
Systems for answering questions via charts~\cite{masry2022chartqa} are typically viewed as black boxes that generate answers directly from a given chart and natural-language question. 
In contrast, \name introduces a glass-box approach that answers questions through a step-by-step reasoning process. This method enhances the alignment between human and machine attention~\cite{sood2023multimodal}.
We anticipate that this approach could lead to significant improvements in chart question answering~\cite{masry2022chartqa} and greater compatibility with explainable AI systems.
}

\subsection{\rv{Extending the Model beyond Bar Charts}}

\rv{
Our modeling approach can be extended to many visualization types besides bar charts.
We analyzed the visualization taxonomy outlined in prior work~\cite{borkin2013makes, borkin2015beyond}, including area, circle, diagram, distribution, grid, line, map, point, table, text, tree, and network, then categorize these visualization techniques into two groups: those that are feasible to extend with minor changes and those that are out of reach, requiring additional features.
}

\begin{figure}[!t]
    \centering
    \subfigure[\rv{An \textit{RV} task with a line chart: ``What was the revenue from newspaper advertising in 1980?''}]{\label{fig:a}\includegraphics[width=0.48\textwidth]{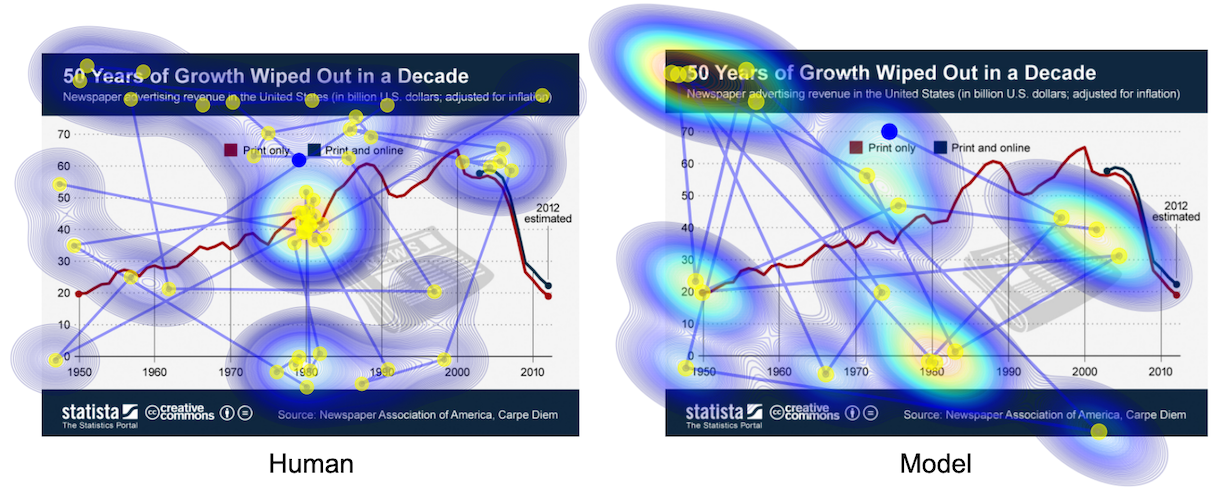}}
    \hspace{0.02\columnwidth}
    \subfigure[\rv{An \textit{F} task with a scatterplot: ``In which countries do people anticipate spending about \$700 for personal Christmas gifts?''}]{\label{fig:b}\includegraphics[width=0.48\textwidth]{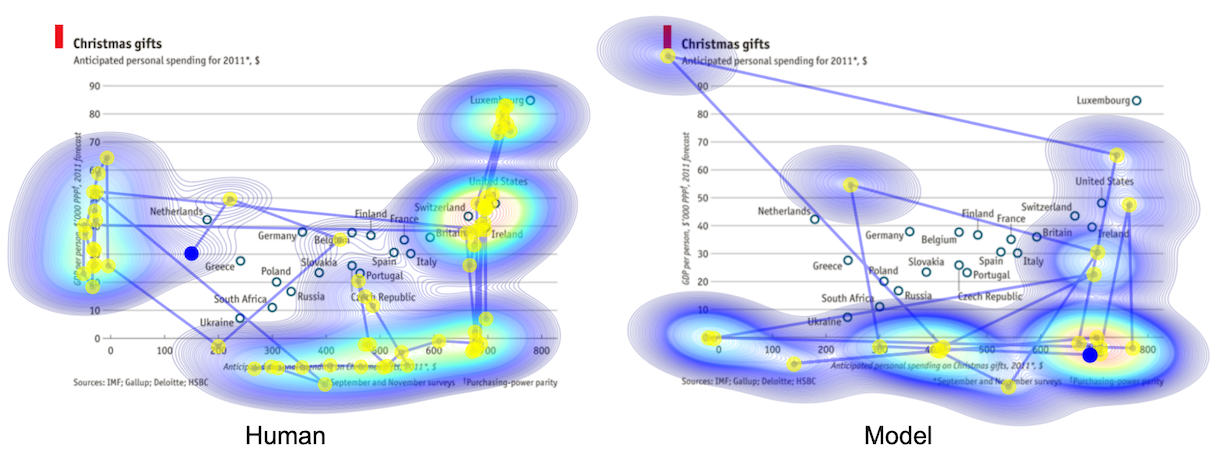}}
    \caption{\rv{Two cases that illustrate the generalizability of the modeling approach, showing the extension of \name to a line chart and a scatterplot. The model's predictions are spatially similar to human ground-truth scanpaths.}}
    \label{fig:case}
    % \vspace{-5mm}
\end{figure}

\rv{
Our modeling approach can be applied to most statistical charts either directly or upon rectification of minor issues. For instance, extending the model to interpret \textit{line charts} and \textit{area charts} is feasible when the axis labels are clearly defined. The trend patterns of lines and areas can be perceived by the peripheral vision as visual guidance.
For \textit{point charts}, such as scatterplots, the model performs well in conditions of sparse data points. However, individual points may be obscured in dense scatterplots, making it difficult to label data when points are cluttered or overlapping. 
\textit{Distribution charts}, such as histograms, and \textit{circle charts}, such as pie charts, are similar to bar charts in that they use the area of marks to represent values. Retrieving exact values from these two presentation types can be imprecise on account of the ranges of the bins and inaccuracies in estimating angles or arc lengths.
Reading \textit{grid charts} (e.g., heatmaps) too is feasible; however, identifying the values necessitates understanding color intensity, a factor that can sometimes lead to ambiguity.
Modeling scanpaths on \textit{tables} or \textit{text} for retrieval tasks is tractable under the current modeling approach, but a lack of visual pattern recognition may render the results poor.
To further examine the generalizability of this category, we considered two additional cases, using a line chart and a scatterplot. We manually labeled the charts, trained the model, and made predictions. As Figure~\ref{fig:case} attests, the trained model performs well for these two chart types when compared to human ground-truth scanpaths.
}

\rv{
Other, sophisticated visualization types are out of reach because they require additional features, particularly prior knowledge and advanced reasoning abilities. For instance, reading \textit{maps} involves associating spatial regions with colors, sizes, or symbols to retrieve related values. Also, when interpreting maps, people rely heavily on preexisting geographical knowledge as a basis for efficient visual searches. Complex designs with intricate structures, such as \textit{diagrams}, \textit{trees}, and \textit{network graphs}, typically require advanced reasoning based on connections. All these skill requirements point to a need for further study in this area.
}

\subsection{\rv{Paths toward Sophisticated Tasks in Chart Question Answering}}

\rv{
Although the model focuses primarily on gaze prediction, it is worth exploring potential improvements for enriching its sophisticated question answering capabilities. We also discuss its limitations.
}

\rv{
Our current model does not achieve the same level of accuracy as the state-of-the-art models represented by the ChartQA benchmark~\cite{masry2022chartqa}. 
Unlike other models that can access the full chart image, \name is limited by its foveal vision and restricted spatial reasoning abilities. For instance, if a bar's height falls between two labeled values, such as 10 and 15, the model might choose either 10 or 15 as its answer when interpreting the axis, failing to provide a more precise value. 
This limitation stems from the constrained spatial perception capabilities of LLMs, which are central to cognitive control.
One possible solution is integrating multi-modal LLMs~\cite{cuarbune2024chart}, for which recent research has demonstrated an accuracy rate of 81.3\%.
}

\rv{
The sense-making process for complex visualizations may be inherently challenging. Even humans often struggle with understanding how the data are encoded, recognizing a given chart's purpose, tackling readability issues, performing numerical calculations, identifying relationships among data points, and navigating the spatial arrangement of graphical elements~\cite{rezaie2024struggles}.
Our model is designed to be straightforward and objective, focusing on analysis tasks related to statistical charts, but it does not fully capture the complexities of visualizations.
A possible enhancement in this respect would be to integrate the model with human sense-making practices~\cite{rezaie2024struggles} or to incorporate a framework of human understanding~\cite{albers2014task}. Such integration could facilitate better simulation of a human-like problem-solving process.
}

\section{Conclusion}

\rv{
Our work contributes a computational model \name that simulates the eye movements on charts when humans solve visual analytical tasks.
The model benefits greatly from its hierarchical gaze control architecture wherein the high-level cognitive controller performs reasoning using memory while the low-level oculomotor controller directs the gaze within visual constraints.  
By following the principle of computational rationality, we are able to train the model in a controlled environment instead of relying on human eye tracking data. This circumvents the costly and time-consuming process of gathering such data. 
The results indicate that the model is better than baselines at generating human-like eye movements during analytical tasks. The predicted scanpaths closely match the spatial positions and temporal order of human scanpaths. While there are limitations to its generalizability and accuracy in question answering, it paves the way for further advances in modeling-based approaches.
}

% \begin{acks}
% \end{acks}

\begin{acks}
% This work was supported by the Research Council of Finland (flagship program: Finnish Center for Artificial Intelligence, FCAI, grants 328400, 345604, 341763; Human Automata, grant 328813; Subjective Functions, grant 357578), the ERC AdG project Artificial User (101141916), and ELLIS Mobility Grant. 
This work was supported by the Research Council of Finland project Subjective Functions (grant 357578), 
Finnish Center for Artificial Intelligence (grants 328400, 345604, 341763),
European Research Council Advanced Grant (no. 101141916),
the Department of Information and Communications Engineering at Aalto University, 
and ELLIS Mobility Grant. 
Y. Wang was funded by the Deutsche Forschungsgemeinschaft~(DFG, German Research Foundation)~-~Project-ID 251654672~-~TRR~161. 
Y. Bai was supported by NUS research scholarship and ORIA program.
A. Bulling was funded by the European Research Council (ERC; grant agreement 801708).
\end{acks}

\bibliographystyle{ACM-Reference-Format}
\bibliography{reference}

% \appendix









\end{document}